\documentclass[twocolumn]{emulateapj}

\shorttitle{The Silicate Emission Zone}
\shortauthors{Kessler-Silacci et al.}

\begin{document}


\title{Probing Protoplanetary Disks with Silicate Emission: \\
    Where is the silicate emission zone?}

\author{J.~E. Kessler-Silacci\altaffilmark{1},
        C.~P. Dullemond\altaffilmark{2},
        J.-C. Augereau\altaffilmark{3},
	B. Mer{\'{\i}}n\altaffilmark{4}, \\
        V.~C. Geers\altaffilmark{4},  
        E.~F. van Dishoeck\altaffilmark{4},
        N.~J. Evans, II\altaffilmark{1},
        G.~A. Blake\altaffilmark{5}, 
        J. Brown\altaffilmark{6} 
}

\altaffiltext{1}{Department of Astronomy, University of Texas at
Austin, 1 University Station C1400, Austin, TX 78712-0259, USA (jes@astro.as.utexas.edu, nje@astro.as.utexas.edu)}
\altaffiltext{2}{Max-Planck-Institut for Astronomy, Koenigstuhl 17, 69117 Heidelberg, Germany (dullemon@mpia.de)}
\altaffiltext{3}{Laboratoire d'Astrophysique de Grenoble, Universit\'e  
Joseph Fourier, B.P. 53, 38041 Grenoble Cedex 9, France (augereau@obs.ujf-grenoble.fr)}
\altaffiltext{4}{Leiden Observatory, PO Box 9513, 2300 RA Leiden, the
Netherlands (vcgeers@strw.leidenuniv.nl, merin@strw.leidenuniv.nl, ewine@strw.leidenuniv.nl)} 
\altaffiltext{5}{Division of GPS, Mail Code 150-21, California
Institute of Technology, Pasadena, CA 91125, USA (gab@gps.caltech.edu)}
\altaffiltext{6}{Division of PMA, Mail Code 105-24, California
Institute of Technology, Pasadena, CA 91125, USA (jmb@phobos.caltech.edu)}

\begin{abstract} 
Recent results indicate that the grain size and crystallinity inferred from observations of silicate features may be correlated with spectral type of the central star and/or disk geometry.  In this paper, we show that grain size, as probed by the 10~$\mu$m silicate feature peak-to-continuum and 11.3-to-9.8~$\mu$m flux ratios, is inversely proportional to $\log {L_{\star}}$.
These trends can be understood using a simple two-layer disk model for passive irradiated flaring disks, {\tt CGPLUS}. 
We find that the radius, $R_{10}$, of the 10~$\mu$m silicate emission zone in the disk goes as $({L}_{\star}/L_{\sun})^{0.56}$, with slight variations depending on disk geometry (flaring angle, inner disk radius).  
The observed correlations, combined with simulated emission spectra of olivine and pyroxene mixtures, imply a grain size dependence on luminosity.  Combined with the fact that $R_{10}$ is smaller for less luminous stars, this implies that the apparent grain size of the emitting dust is larger for low-luminosity
sources. In contrast, our models suggest that the crystallinity is
only marginally affected, because for increasing luminosity, the zone
for thermal annealing (assumed to be at $T>800$ K) is enlarged by
roughly the same factor as the silicate emission zone. 
The observed crystallinity is affected by disk geometry, 
however, with increased crystallinity in flat disks. 
The apparent crystallinity may also increase with grain growth due to a corresponding increase in contrast between crystalline and amorphous silicate emission bands.  
\end{abstract}





\section{Introduction}

Silicate emission features from circumstellar disks have now been observed toward large samples of young stars, ranging from sub-stellar mass brown dwarfs to Herbig Ae/Be (hereafter HAEBE) stars up to 10 times the mass of the sun  \citep[see, e.g.,][]{bouwman_01, vanboekel_03, vanboekel_05, me_05, me_06, apai_05}.  The most commonly observed silicates are amorphous olivines and pyroxenes and their crystalline, Mg-rich forms, enstatite (MgSiO$_3$) and forsterite (Mg$_2$SiO$_4$).  
For amorphous olivine and pyroxene, the features due to distinct stretching or bending modes of the silicon-oxygen bonds are blended into two broad features near 9.8~$\mu$m (stretch) and 18~$\mu$m (bend).  In addition to providing information about the grain composition and crystallinity, 
silicate features have been used as a probe of the grain size \citep{vanboekel_03, me_06}. 
Since the silicate emission features arise from the optically thin surface layer of the disks, and larger grains are expected to settle to the disk midplane, the grain size and crystallinity indicated by these features applies primarily to the smallest grains in the disk.

Studies of silicate emission features around HAEBE stars and comets \citep{wooden_02}, indicated that mature disks and comets had silicate emission features characteristic of up to 30\% crystalline silicates, while much lower crystallinity was observed toward many young disks \citep{bouwman_01}, embedded objects \citep{bowey_98, whittet_97} and through the ISM toward the galactic center \citep{kemper_04, kemper_05}.  This suggests an evolutionary trend in which amorphous silicates from the ISM are crystallized {\em within} circumstellar disks.
Indeed, the  types of silicate features observed seem to support this scenario.    Van Boekel et al.~(2005) found that sources showing large mass fractions of crystalline silicates also possessed higher mass fractions of large grains, suggesting that crystallinity is somehow related to grain coagulation in disks -- this is not necessarily a causal relationship.  For disks around sub-stellar mass stars, the spectra of flat disks, in which dust coagulation and settling have resulted in smaller flaring angles, often show higher crystallinity and larger grain sizes than those of more flared disks \citep{apai_05}.  Additionally, \citet{me_06} show that the grain size indicated by the 10~$\mu$m feature is larger for disks around M-stars than for disks around A/B-stars, which they interpret as an indication that the silicate feature probes different regions of the disk depending on stellar luminosity. 

In order to examine the physical processes responsible for these trends, we use a 2-layer disk model to trace the silicate emission zones for a range of stellar masses from 0.003--7 M$_{\odot}$.  The relationship between observed silicate emission feature characteristics and stellar luminosity is described in $\S$2. The disk models are described in $\S$3.  Finally, in $\S$4, the results and implications are discussed.

\section{Silicate Emission Features, Grain Size and Stellar Luminosity}

\begin{figure}[h]
\includegraphics[angle=90,scale=0.46]{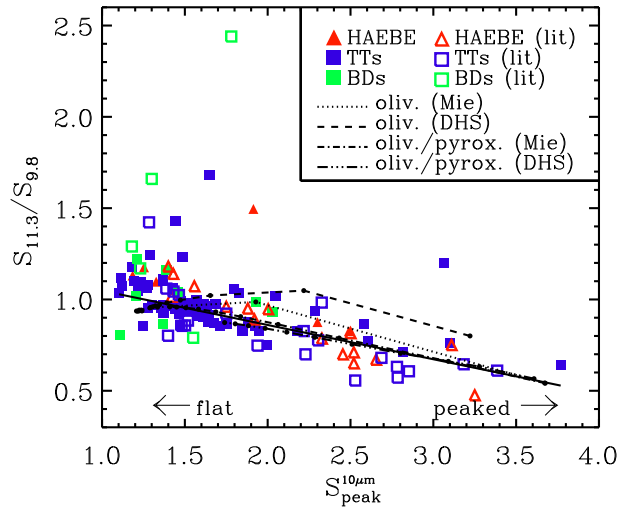}
\caption{Silicate 10~$\mu$m feature strength vs.\ shape. Green squares are brown dwarfs with $M_{\star}<75$ $M_{Jup}$.  Peaked silicate features show up toward the right of the plot and flat features show up toward the left. Blue squares are low-mass stars with $M_{\star}>75$ $M_{Jup}$.  Red triangles are HAEBE stars.  Filled symbols are from the Spitzer c2d sample. Open symbols are from \citet{vanboekel_03, przygodda_03, me_05, apai_05}. The black line denotes a fit to the points with $y<1.1$, $y= -(0.18{\pm}0.02)x + (1.23{\pm}0.03)$, with $r=-0.77$, indicating a probability of $>$99.05\% that x and y are correlated. Models of pyroxene and olivine mixtures, treating the dust as filled homogeneous spheres (Mie) or distributions of hollow spheres (DHS) with grain sizes from 0.1--20~$\mu$m, from \citet{me_06} are overlaid as dashed/dotted lines. \label{fig:vb10all}}
\end{figure}

The connection between silicate emission strength and shape was first noted by \citet{vanboekel_03} for a sample of HAEBE stars, and has now been seen for a number of sources, including HAEBE stars, T Tauri stars and, more recently, a small sample of brown dwarfs \citep{przygodda_03,me_05,me_06,apai_05}.  These results are shown in Figure~\ref{fig:vb10all}, with the addition of $\sim$40 new sources more recently observed by Olofsson et al. (in prep.) as part of the ``From Molecular Cores to Planet-Forming Disks  (c2d)'' Spitzer Legacy program \citep{evans_03}, including 9 brown dwarf candidates selected from \citet{allers_06}. As was shown in \citet{vanboekel_03} and \citet{me_05}, the flattening of the observed shape, indicated by the 11.3-to-9.8~$\mu$m flux ($S_{11.3}/S_{9.8}$), and decrease in the observed strength ($S^{10{\mu}m}_{peak}$) of the silicate features can be tied to increases in the amorphous olivine/pyroxene grain sizes, such that grain size increases from the bottom right to the top left of Figure~\ref{fig:vb10all}.  Crystalline forsterite and/or PAH emission can result in increased flux near 11.3~$\mu$m and therefore cause deviation from this trend \citep[see also][]{me_06}.  In particular, spectra with $(S_{11.3}/S_{9.8})/S^{10{\mu}m}_{peak}>1.0$--1.1 (approximately 1/5$^{th}$--1/3$^{rd}$ of the sample shown in Figure~\ref{fig:vb10all}) cannot be explained with models of pure amorphous silicates. Overlaid in Figure~\ref{fig:vb10all} are parameters calculated from modeled spectra of pure olivine and olivine-pyroxene (75\%:25\%) mixtures for filled homogeneous spheres (using Mie theory) or distribution of hollow spheres (DHS \citep{min_05}, representing porous grains).  The olivine-pyroxene mixtures correspond well with the observed trend. 

\begin{figure*}
\begin{center}
\includegraphics[angle=90,scale=0.64]{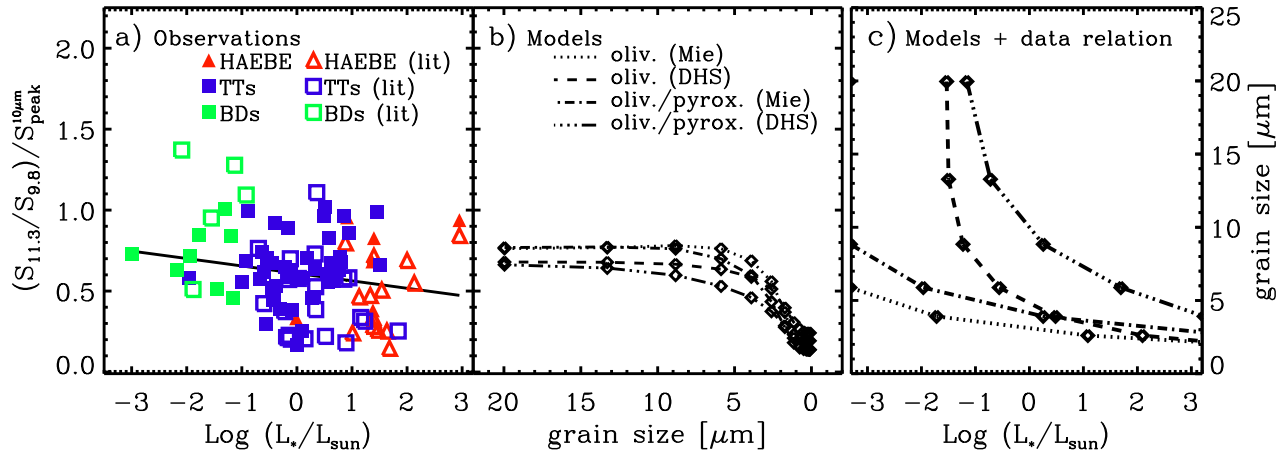}
\caption{$(a)$ Grain size parameter vs.\ stellar luminosity. Symbols are as in Figure~\ref{fig:vb10all}. The black line denotes the correlation, $y=-(0.05{\pm}0.02) x + (0.61{\pm}0.02)$, with $r=-0.23$, indicating a probability of $>$97\% that x and y are correlated. $(b)$ Models of grain mixtures (as in Figure~\ref{fig:vb10all}) relating the grain size to $(S_{11.3}/S_{9.8})/S^{10{\mu}m}_{peak}$. $(c)$ Models relating $\log (L_{\star}/L_{\sun})$ to grain size, using the relation derived from $(a)$ applied to the models in $(b)$. Line types are the same as in $(b)$. \label{fig:vb10lum}}
\end{center}
\end{figure*}

Figure~\ref{fig:vb10all} shows a trend in which the disks around the lowest luminosity brown dwarfs have larger grain sizes than those around T Tauri or HAEBE stars. This is made more apparent in Figure~\ref{fig:vb10lum}a, where we plot the grain size parameter versus the logarithm of the stellar luminosity for the same sample as shown in Figure~\ref{fig:vb10all}.  Indeed, although the exact relationship between grain size and $(S_{11.3}/S_{9.8})/S^{10{\mu}m}_{peak}$ may vary depending on the grain composition/shape (Figure~\ref{fig:vb10lum}b), there appears to be an inverse correlation between the grain size parameter and the stellar luminosity.  Brown dwarfs have larger $(S_{11.3}/S_{9.8})/S^{10{\mu}m}_{peak}$ than T Tauri stars, which have larger $(S_{11.3}/S_{9.8})/S^{10{\mu}m}_{peak}$ than HAEBE stars, and although there is a large degree of scatter, the correlation is significant, with a $>$97\% probability that x and y are correlated and all data lying within 2$\sigma$ of the fitted trend. \citet{me_06} postulated that the observed relationship between grain size and spectral type could be explained if the silicate features probed different distances from the star, probing smaller grains located farther out in the disks around more luminous stars.  

Using the relation between $(S_{11.3}/S_{9.8})/S^{10{\mu}m}_{peak}$ and $\log L_{\star}$ from observations in Figure~\ref{fig:vb10lum}a and the $(S_{11.3}/S_{9.8})/S^{10{\mu}m}_{peak}$ and grain size from modeled spectra in Figure~\ref{fig:vb10lum}b, one can determine a relationship between grain size and luminosity, shown in Figure~\ref{fig:vb10lum}c. The feature strength and shape for a particular grain size can be quite different for different silicate compositions/shapes (Figure~\ref{fig:vb10lum}b,\citealp{me_06}). This in turn affects the precise relationship between grain size and stellar luminosity (Figure~\ref{fig:vb10lum}c).  In particular, the DHS models show stronger features than the Mie models for the same grain size, and thus for low luminosity stars, the DHS models require grain sizes that are much larger ($\sim$20~$\mu$m) to produce the weak features.  Nevertheless, the conclusion that apparent grain size is inversely proportional to luminosity is robust for all reasonable grain models.

\section{Models of the Silicate Emission Zone}

In order to understand possible physical explanations for the correlation shown in Figure~\ref{fig:vb10lum}a, we model the disks around a set of stars ranging from $\sim$$10^{-4}$--$10^4$ $L_{\odot}$ to better identify how the silicate emission zone --- the portion of the disk contributing to the 10~$\mu$m feature --- changes as a function of the stellar luminosity.   
We use the 2-layer model of \citet{dullemond_01} ({\tt CGPLUS}), which is based on work by \citet{CG97} and \citet{chiang_v01}.  This model assumes vertical hydrostatic equilibrium, resulting in a flared disk structure. 
Directly illuminated dust grains in the optically thin surface layer re-radiate half of the energy absorbed from the central star down into the disk interior and thus regulate the interior disk temperature.  The other half is emitted away from the disk and can be observed as optically thin emission from the disk (e.g., emission features). 
We employ a new version of CGPLUS in which the originally vertical blackbody puffed-up inner dust wall model is replaced by a smoothly rising disk height from $H=0$ at the dust evaporation radius to $H=\sqrt{kT_{mid}R^3/{\mu} m_p G M_{\star}}$ at a $\sim$20\% larger radius, where $T_{mid}$ is the midplane temperature at radius $R$, ${\mu}$ is the gas mean molecular weight, and $m_p$ is the mass of a proton (see \citealp{calvet_02} for a description of this
method to mimic a rounded-off inner dust rim).
The location of the inner rim of the disk is determined by the dust sublimation temperature (typically 900-1600 K for silicates, assumed here to be 1500 K as found by \citealp{duschl_96}). The inner radius is set to $R_{\star}$ if the dust sublimation temperature is not reached. 
 Opacities used in {\tt CGPLUS} are from \citet{laor93}, for mixtures of 0.1 $\mu$m spherical amorphous olivine and crystalline graphite grains.

\begin{figure*}[t]
\begin{center}
\includegraphics[angle=90,scale=0.54]{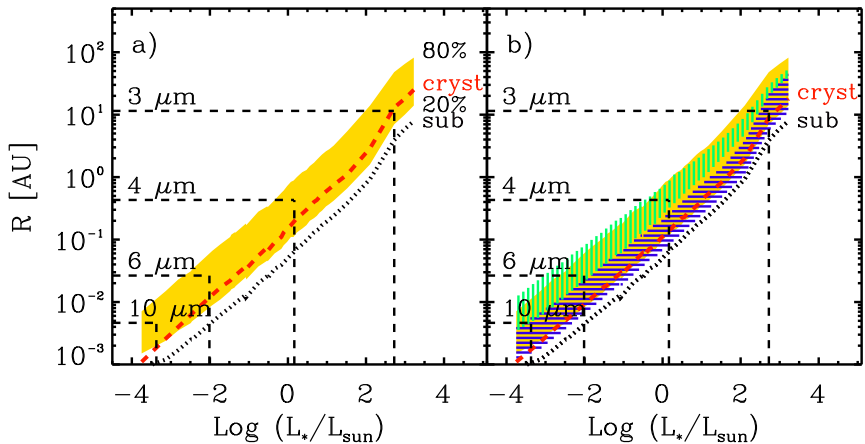}
\caption{Radius probed by 10~$\mu$m feature as a function of stellar luminosity.   $(a)$ The yellow solid region shows the radii contributing 20--80\% of the 10~$\mu$m emission for flared disks.   $(b)$ The additional blue horizontal striped and green vertical striped regions show the same for flat disks and flared disks with inner radii 3 times larger than the dust sublimation radius, respectively. In both panels, the black dotted and red dashed lines show the radii corresponding to $T=1500$ K and $T=800$ K, above which temperature silicates will sublimate and crystallize via annealing, respectively.  The vertical black dashed lines show the luminosities corresponding to grain sizes of 3, 4, 6, and 10~$\mu$m for the grain size vs.\ stellar luminosity trend in Figure~\ref{fig:vb10lum}c. The horizontal black dashed lines show the radii corresponding to those luminosities, using the relationship shown by the filled regions, $\log(R)=-0.45+0.56\log(L_{\star}/L_{\sun})$. \label{fig:2by2}}
\end{center}
\end{figure*}

To enable comparison between the models, the disk mass is assumed to be a simple function of the stellar mass, $M_{disk}=0.03$ $M_{\star}$. Using $M_{disk}{\propto}M_{\star}^2$ might be a better assumption \citep{hueso_guillot_05,dullemond_06_coredisks}, but has a very small effect (an offset of $<0.02$ AU) on the radius of the silicate emission zone. Stellar luminosities, masses and surface gravities are related using the pre-main-sequence evolutionary tracks of \citet{baraffe_98,baraffe_03} for $L_{\star}=0.0002$--0.5~$L_{\sun}$ and \citet{siess_00} for $L_{\star}=0.1$--$2000$~$L_{\sun}$. To check consistency, both models are used from $L_{\star}=0.1$--0.5~$L_{\sun}$, and we find no discontinuities in the overlap region (see Figure~\ref{fig:2by2}). The outer radius of the disk corresponds to where gas would no longer be bound to the star ($0.2 r_g$, where $r_g = GM_{\star}{\mu}m_p/kT$; see, e.g., \citealp{dullemond_05_ppv}).  The disk surface density is described as $\Sigma = \Sigma_0 (R / 1\, \rm{AU})^{\beta}$, where $R$ is the radial distance from the star, $\Sigma_0$ is a constant set by the disk mass, and $\beta=-1$ for all disks.  Due to the fact that we extend the disk out to the radius where the gas becomes unbound, the disk in these outer regions will be very geometrically thick. This makes it difficult to explore inclination effects using the simple two-layer model, and true 2-D/3-D disk models are beyond the scope of this paper. Instead we always assume that the disk is face-on ($i=0^{\circ}$) in our models.
However, variations in the inclination angle could contribute to the scatter shown in Figure~\ref{fig:vb10lum}a, with higher inclinations resulting in decreased 10~$\mu$m feature strength. 

{\tt CGPLUS} produces both spectral energy distributions and images at each wavelength for each model.  In order to evaluate the 10~$\mu$m emission zone, we calculate the radially integrated cumulative flux of the 10~$\mu$m emission image.  The 10~$\mu$m emission zone is then defined as the radial zone where the cumulative flux reaches 20\%--80\% of the total 10~$\mu$m emission. Figure~\ref{fig:2by2}a, shows the 10~$\mu$m emission zone (yellow solid region) for the resulting models.  The location of the zone increases roughly as $R_{10}=0.35 \,{\rm AU}\, (L_{\star}/L_{\sun})^{0.56}$.  This relationship can be easily approximated under the assumption of equilibrium between the absorbed and emitted energy \citep[e.g.,][equations 1 and 2]{backman_paresce_93}, with the simplifying assumptions that $1)$ the emission is coming from an optically thin region of the disk, $2)$ only a narrow range of temperatures is probed at 10~$\mu$m, and $3)$ that the grains absorb to first order as blackbodies.  The resulting distance between the illuminating star and the emitting grains is $R \propto L^{0.5}$, which is very similar to the relation found here.

The silicate emission therefore arises from very different regions in disks around brown dwarfs ($R_{10}\le0.001-0.1$ AU) than in disks around HAEBE stars ($R_{10}\ge0.5-50$ AU).  It is likely that the conditions in these disks, and thus properties of the silicates, are very different at these radii.  
We will now use some simple approximations to qualitatively indicate how radial conditions can affect the silicate properties (grain size, crystallinity) that are probed with the 10~$\mu$m features. 

\subsection{Effects of Radial Conditions on Grain Size}

Figure~\ref{fig:vb10lum}a shows that the grain sizes interpreted from observations of the 10~$\mu$m feature appear to be related to the stellar luminosity \citep[see also][]{me_06}.  To simplify, we will attempt to explain the observed trends using {\em Ansatz} that the grain size distribution with radius is the same in all disks, and that only the silicate emission region changes. We are aware that this may not be quite realistic, but coagulation is at present insufficiently understood \citep[see, e.g.,][for review]{dominik_05} and we therefore wish to study -- in isolation -- the effect of the effect of the radial location of the silicate emission zone on the observed grain sizes.

The grain size distribution vs.\ luminosity deduced from observations and shown in  Figure~\ref{fig:vb10lum}c, has been used to infer the effect of radial conditions on grain size.   The relation for DHS olivene-pyroxene mixture has been adopted since this model best represents the trend shown in Figure~\ref{fig:vb10all}. Using the modeled variation, the grain sizes of 3, 4, 6 and 10~$\mu$m would correspond to the luminosities shown as dashed vertical lines in Figure~\ref{fig:2by2}. Using the modeled variation of the location of the silicate emission zone with stellar luminosity shown by the filled regions, we then calculate the radii corresponding to those luminosities (shown as dashed horizontal lines).  This results in grain sizes that decrease as a function of disk radius, which seems reasonable since coagulation and dust settling processes are believed to occur more quickly at smaller radii \citep[see, e.g.][]{dullemond_05}. 
Thus this ``reverse modeling'' shows that the observation that later type stars appear to have larger grains (as deduced from their spectra) might be due, at least in part, to the fact that grains are expected to be  bigger at smaller radial position in the disk.

The example presented above is indicative of how stellar luminosity, and the corresponding location of the 10 $\mu$m silicate emission zone, might affect the grain size deduced from silicate emission features.  
However simple, this example illustrates that the location of the silicate emission zone is dependent on stellar luminosity.  This effect cannot be ignored when one seeks to interpret the grain sizes derived from silicate emission features in terms of grain growth in the disk as a whole.

\subsection{Effects of Radial Conditions on Crystallinity}

Although crystalline silicates have been observed in disks around HAEBE stars, T Tauri stars, and brown dwarfs, the mechanism for conversion from the primarily amorphous ISM silicates 
is not completely understood. 
Spatially resolved mid-IR interferometry of disks around HAEBE stars \citep{vanboekel_04_nature} indicate that the crystallinity is much higher at small radii ($<2$ AU) than in the outer disk.  This result supports the theory of crystallization via annealing at high temperatures in the inner disk, followed by transport to larger radii.   In this case, the amount of crystalline silicates present in the 10~$\mu$m emission zone is a function of both the radial location of the zone and the timescale for radial mixing.

To understand the effects of probing different radii on the observed 10~$\mu$m crystallinity, we plot in Figure~\ref{fig:2by2}a the radii corresponding to $T_s=800$ K, the minimum temperature required for annealing amorphous olivine \citep{duschl_96,gail_98}.  If crystallization is occuring primarily through annealing, then we would expect dust to be crystallized at radii equal to or smaller than the dashed red line in Figure~\ref{fig:2by2}a.  
As the 10 $\mu$m emission arises from the disk surface layer, we use only the 
temperature of the surface layer to calculate the region.
For simplicity, the default version of the {\tt CGPLUS} model, with a single
small grain size throughout the disk, is used in these calculations.

Although the 10~$\mu$m emission zone moves to larger radii for HAEBE stars with respect to brown dwarfs, the crystallization zone also expands to include larger radii; more luminous stars can heat dust farther out in the disk.  
Thus, we expect the percentage of crystalline silicates emitting at 10~$\mu$m to be nearly independent of the luminosity of the central object.  
There does appear to be a small increase in $R_{cryst}/R_{10}$ with increasing stellar luminosity, however, such that we might expect to observe more crystalline silicates from HAEBE star disks than brown dwarf disks.  
In contrast, observations show that crystallinities can be quite large in disks around brown dwarfs (\citealp{apai_05}; {Mer{\'{\i}}n} {et~al.} 2006, submitted).  This difference could be explained by accretion, which would result in viscous heating of the midplane of the inner disk, thus leading to additional crystallization at smaller radii. Additionally, the degree of crystallinity in these disks could be affected by the presence of additional crystallization methods or the effectiveness of radial mixing, which are not considered here.
Overall, the model  predicts nearly constant crystallinity as a function of luminosity.

\subsection{Effects of Disk Geometry}

Because disk geometry affects the incidence of stellar light onto the surface of the disk, it should also affect the absorption and resulting emission of this light via the 10~$\mu$m feature.
The effects of disk geometry are shown in Figure~\ref{fig:2by2}b.  The yellow solid region shows the 10~$\mu$m emission zone for models of flared disks, as in Figure~\ref{fig:2by2}a.  

The blue horizontal striped region shows the 10~$\mu$m emission zone as calculated for models of flat disks, possibly resulting from dust settling, with otherwise the same parameters as the flared disks.  
The flaring angle of the disk determines the angle at which the stellar radiation penetrates the disk surface. Since the radiation only penetrates to $\tau=1$ (along the line of sight), the true (vertical) depth decreases with the incidence angle $\phi$ as $\sin(\phi)$. So for a flat disk (with a steep drop-off of $\phi$ with radius), the depth at which the stellar radiation heats the disk decreases very steeply with radius.  This means that the amount of matter contributing to the 10~$\mu$m feature also decreases much faster as a function of radius for a flat disk than for a flared disk.  So the 10~$\mu$m feature from a flat disk is dominated by emission from shorter radii, while the feature from a flared disk includes also emission that is spread out over a larger portion of the disk. 
With our prescriptions for grain size and crystallinity, this means that the 10~$\mu$m emission from flat disks would probe predominantly smaller radii, with larger and slightly more crystalline silicates, as seen in \citet{apai_05}.

The green vertical striped region in Figure~\ref{fig:2by2}b shows the 10~$\mu$m emission zone for models of flared disks with inner radii that are larger than the dust sublimation radius.  By moving the inner radius of the disk outward, we simulate the effect of clearing the inner disk, e.g., by a planet or photoevaporation.  Models of this type successfully reproduce spectral energy distributions observed for several T Tauri star disks, such as GM Aur, DM Tau and CoKu Tau 4 \citep{calvet_05,forrest_04,dalessio_05}. 
As seen in Figure~\ref{fig:2by2}b, the addition of an inner hole in these models means that the 10~$\mu$m emission is now probing larger radii (especially for low luminosity stars). 

As shown in Figure~\ref{fig:2by2}b, the disk geometry (flaring angle, inner holes) can affect the radial location of the 10~$\mu$m emission zone. 
The presence of a variety of disk geometries could contribute to the observed scatter in the trend noted in Figure~\ref{fig:vb10lum}, while preserving the overall correlation with $L_{\star}$, as observed.  

\section{Discussion and Implications}

In this paper we have demonstrated that the region of a protoplanetary disk that is probed by 10~$\mu$m silicate emission features is related to the luminosity of the central object; specifically, the 10~$\mu$m emission zone is at larger radii for disks around more luminous stars, $R_{10}=0.35 \,{\rm AU}\, (L_{\star}/L_{\sun})^{0.56}$.
As a result, the grain sizes inferred from observations of silicate emission features may reflect the location of the silicate emission zone more than the properties of the bulk silicate dust in the observed disk.
If the grain size is a function of the radial position in the disk (as shown here by extrapolating from the observed feature shape/strength vs.\ luminosity and the modeled grain size vs.\ shape/strength relations), then the correlation of grain size with stellar type seen in the sample of \citet{me_06} and the generally large grain sizes inferred from observations of dust around very low mass stars by \citet{apai_05} may be related directly to the location of the 10~$\mu$m emission zone in these disks.  Additionally, the 10~$\mu$m emission zone can vary with disk geometry (flaring angle, inner disk radius), which may result in deviations from the trend for individual objects. 

The amount of crystalline silicates contributing to the observed 10~$\mu$m feature, if dependent primarily on the disk temperature, would {\it not} be strongly affected by stellar luminosity, as both the crystallization region and the 10~$\mu$m emission zone increase in size with increasing $L_{\star}$. However, if the grain size is a function of radial position in the disk, then 
the apparent crystallinity inferred from the 10~$\mu$m feature may be affected, due to the fact that the contribution to the spectrum from amorphous silicates diminishes more strongly with grain growth (0.1--10~$\mu$m) than does that from crystalline silicates \citep[c.f.,][Figures 6--8]{me_06}.  This effect 
may contribute to the large crystallinities seen toward brown dwarfs (\citealp{apai_05}; {Mer{\'{\i}}n} {et~al.} 2006, submitted) and the restriction of significant crystalline mass fractions ($>$15\%) to those HAEBE disks that also possess $>$80\% big (1.5~$\mu$m) grains by mass \citep{vanboekel_05}.

The relationship of the location of the 10~$\mu$m emission zone with stellar luminosity and disk geometry demonstrated here will also affect studies of correlations of silicate properties (as deduced from silicate emission features) with stellar parameters, age, and disk evolution.  In particular, relationships of grain properties with disk/stellar evolution or age are difficult as the stellar luminosity and disk geometry are expected to change with evolutionary state of the system and both are shown here to affect the location of the 10~$\mu$m emission zone.  This study indicates the necessity for careful modeling of disks in order to deconvolve the effects of variations in the 10~$\mu$m emission zone and true changes in silicate dust properties. 






\acknowledgments

Support for this work was provided through Contract Numbers 1256316, 
1224608 and 1230780 issued by the Jet Propulsion Laboratory, California
Institute of Technology under NASA contract 1407.  
Astrochemistry in Leiden is supported by a NWO Spinoza 
and NOVA grant, and by the European Research Training Network "The Origin of
Planetary Systems" (PLANETS, contract number HPRN-CT-2002-00308). The
authors would like to thank the ApJ referee for many helpful comments and suggestions.

\bibliographystyle{apj}

\clearpage

\clearpage

\clearpage

\end{document}